\documentclass[a4paper,9pt]{article}
\usepackage{ulem}
\usepackage{color,amsmath}
\usepackage{amsfonts}
\usepackage{amssymb}
\usepackage{graphicx}
\usepackage{epsfig}
\usepackage{enumerate}
\usepackage{stmaryrd}
\usepackage{varioref,exscale,relsize}
\usepackage{amsthm}
\usepackage{mathrsfs}
\usepackage{amsfonts}
\IfFileExists{url.sty}{\usepackage{url}}
                     {\newcommand{\url}{\text}}

\usepackage{anysize}
\marginsize{2cm}{1cm}{3cm}{0.1cm}

\newtheorem{prop}{Proposition}

\newtheorem{remark}{Remark}
\newtheorem{theo}{Theorem}

\newtheorem{defn}{Definition}

\begin{document}

\begin{center}
\large{\textbf{Feature Selection for Functional Data}}
\end{center}

\begin{center}
  Ricardo Fraiman$^{a}$, Yanina Gimenez$^{b}$ and Marcela Svarc$^{b}$ \\
  $^{a}$ Centro de Matemática, Facultad de Ciencias, Universidad de la Rep\'ublica.\\
  $^{b} $Departamento de Matemática, Universidad de San Andr\'es and CONICET.\\
\end{center}

\section*{Abstract}
We herein introduce a general procedure to capture the relevant information from a  functional data set in relation to a statistical method used to analyze the data, such as, classification, regression or principal components. The aim is to identify  a small subset
of functions  that can ``better explain"  the model, highlighting its  most important features. We obtain consistency results for our proposals. The computational aspects are analyzed, an heuristic stochastic algorithm is introduced and real data sets are studied.

\textit{\textbf{Keywords:} Variable Selection, Classification, Regression, Principal Components.}

\textit{\textbf{MSC[2010]} 62H30,62H25, 62J05}

\section{Introduction}

During the past years technological advances made possible the processing and storing of real time information. Consequently, functional data arose across several fields, such as, finance, meteorology, medicine, among others. Hence, to handle these data new statistical procedures  began to be developed. Classical statistical tools had to be rethought under this framework from a theoretical and empirical approach. In addition,  new problems concerning the nature of the data had to be tackled. There is a vast literature devoted to these topics, the most classical contributions are the books  by Ramsay and Silverman (\cite{RS02},\cite{RS05}) and Ferraty and Vieu \cite{FV06}. More recently,  the books by Ferraty and Romain \cite{FR11} and Horv\'{a}th and Kokoszka \cite{HK12}  establish the state of the art on functional data analysis,  reviewing the advances towards principal components analysis, regression and classification, among other problems.

The problem of feature extraction for multivariate problems has extensively been studied. The methods that show better results usually assess the importance of each variable based on the role that it plays in the statistical procedure being followed (classification, regression, principal components, etc).  The list of strategies available to handle these problems is large, however it is worth to mention that there are at least two proposals that have been studied through several multivariate procedures, namely the Bayesian model averaging (BMA) and
    the ``least absolute shrinkage and selection operator" (LASSO).
    The former approach are  Bayesian proposals introduced by Frayley and Raftery (\cite{FR02}
     and \cite{FR09}). Therein they analyze the problem of unsupervised classification.
     Hoeting et al. \cite{HRM02}, extent those ideas to the linear model. The LASSO
    was proposed by Tibshirani \cite{T96}, who stated that ``It shrinks some
    coefficients and sets others to 0, and hence tries to retain the
    good features of both subset selection and ridge regression...The
    LASSO minimizes the residual sum of squares subject to the sum
    of the absolute value of the coefficients being less than a
    constant. Because of the nature of this constraint it tends to
    produce some coefficients that are exactly 0 and hence gives
    interpretable models". More recently Witten and Tibshirani (\cite{WT10},\cite{WT08}) applied those principles to cluster and principal components. A detailed description of this method can be found in \cite{HTF01}.


The scenario is completely different in the functional data
framework where there is much less literature available. Several
authors address the problem of variable selection in the
regression model. James et al \cite{JWZ09} extend the ideas of
variable selection for regression studying the regression model
when the predictor is functional and the response is scalar. Their
goal is to estimate the regression function smoothly and sparsely,
controlling its derivative and extending the ideas of the LASSO
estimate. They claim that regions where  the regression function
is not null correspond to places where there is a relationship
between the predictor and response variables, alternatively when
it vanishes there is no relationship. They divide the time period
into a grid of points. Then they use variable selection methods to
determine whether the $d$th  derivative of the regression function
is zero or not at each time point. They implicitly assume that the
$d$th derivative will be sparse. Lee et al. \cite{LBP12} also
tackle this problem. They consider a sparse functional linear
regression model which is generated by a finite number of basis
functions in an expansion of the coefficient function. They adopt
the idea of variable selection in the linear regression setting
adding a weighted $L_1$ penalty to the traditional least squares
criterion. Zhou et al \cite{ZWW13} introduce a functional linear
model with zero-value coefficient function at sub-regions
extending the SCAD ideas to this framework. In the three references above mentioned, the procedures described are suitable under different notions of sparsity, an assumption that  may be restrictive. Cuevas
\cite{C14} highlights that functional data can be discretized
where standard variable selection for high dimensional data could
be applied. Aneiros and Vieu \cite{AV14} study the problem of variable selection in the regression setting, when the covariates are functionals. They highlight that  even though, due to the practical treatment of the data that involve discretization, these problems are usually treated as classical high dimensional problems, where $p>>n,$ hence ordinal variable selection procedures that are suitable for this setting are applied. However, they stress the importance of developing new variable selection techniques that take profit of the continuous nature of the variables. Along similar lines, Aneiros et al \cite{AFV15} consider the problem of
variable selection for high dimensional data in partial least
squares, in a very general setup. In particular,
since they work on an abstract setting they allow to incorporate
functional data to the model. Gertheiss et al \cite{GMS13} also consider the problem of variable selection in regression with scalar response and functional covariates, they propose a penalized likelihood approach that combines selection of the functional predictors and
estimation of the smooth effects for the chosen subset of predictors.  McKeague and Sen \cite{MS10} and Kneip et al \cite{KPS15}, deal with the problems of functional regression models assuming the existence of  an unknown number of points on the functional covariates that have special impact on the response. In  \cite{BGSV14} there are several recent results  related to these topics.  For instance,  Aneiros and Vieu \cite{AV14-2} introduce a variable selection procedure for the case of linear and partial linear regression when some covariates are functional, they take into account the natural specifications of the functional variables. Also, Abramowicz et al \cite{AHHPSSV14} present a procedure for performing an ANOVA test for Functional Data that as a byproduct selects intervals where the groups significantly differ. The problem of variable selection for logistic regression has been studied by Matsui \cite{M14}.

The problem of variable selection in classification has also been studied.
Tian and James \cite{TJ13}   introduce an interpretable
dimension reduction procedure to classify functional data. On a first stage,
they reduce the dimension of the data considering projections, then
they proceed to the classification step. More precisely,
let $\left\{X(t), \ t\in [a,b]\right\}$ be a stochastic process
and $Y$ a categorical response variable, without lost of
generality $Y=\{0,1\}$. Let $\left\{f_1,\ldots,f_d\right\}$ be
a set of functions in $[a,b]$, and denote $Z_j$ the projections of $X(t)$ on direction $f_j$, i.e.
$Z_j=\int_a^b X(t) f_j(t) dt \text{ con } \ j=1, \ldots, d.$
They perform the classification procedure in the low dimensional space. Hence, their aim is to retain a subset of functions that minimizes the misclassification rate. The key step of the proposal relays on the choice of the set of functions $\left\{f_1,\ldots,f_d\right\},$ which are selected from the set of constant and linear functions defined on subintervals of $[a,b]$. They also tackle the computational problem describing a competitive stochastic algorithm for choosing $f_j$ for $j=1,\dots,d$.
Also, Martin-Barragan et al \cite{MLR14} present a support vector machine based classifier for functional data that has good classification ability and it is easy to interpret. Fauvel et al \cite{FDZF15} present a nonlinear parsimonious feature selection algorithm for the classification of hyperspectral images and the selection of spectral variables.
%

Our approach is different, it is designed to be used following a ``satisfactory''   analysis of the data,
 by which we mean that the data have previously been subjected to classification,
principal components or some other form of statistical analysis. Additionally, we have a set of functions, $\mathcal{A}=\left\{f_1,\ldots,f_p\right\},$ that provides relevant features of the data. Our aim is to choose a subset of $\mathcal{A}$ that retains almost all the relevant information of the data set related to the statistical procedure that is being studied.
We extend the ideas introduced by Fraiman et al \cite{FJS08}  for several multivariate procedures.  They assume that a statistical analysis of a multivariate data has been  successfully carried out, hence their idea is to   \textit{blind} unnecessary or redundant variables (by means of the conditional expectation), keeping the subsets of variables that better reproduce the output of the statistical procedure in the high dimensional space. Our procedure can be applied to a broad family of statistical problems.  In this paper we extend the \textit{blinding} procedure to the functional framework for the problems of classification, principal components and regression.

The remainder of the paper is organized as follows. We introduce the main definitions in Section 2. In Section 3 we study the classification problem. Section 4 is devoted to principal components analysis and Section 5 to functional regression. Finally, in Section 6 we consider some numerical aspects and analyze several well known data sets. Our concluding remarks are made on Section 7.  All proofs are given in the Appendix.

\section{Main Definitions and notation} \label{main}

We start by fixing some notation that will be used throughout the manuscript. Let $X: \Omega  \rightarrow L^2[a,b]$ be a random element $\{X=:X(t):t \in [a,b] \}$  defined on a rich enough probability space
$(\Omega, \mathcal A, P)$ 
where  $[a,b] \subset \mathbb R$ is a finite interval. As it is
well known, in  practice, the data will be available on a grid $ a\leq s_1 < \ldots < s_N \leq b$ which will be also denoted by $[a,b].$

A population statistical model is given by the function
$\psi(P):=\psi(X,g)$, for $\psi:L^2[a,b] \times \mathbb{G} \to L^2[a,b]$, where $\mathbb{G}$ is a separable metric space. The
output of the statistical procedure is a random function in
$L^2[a,b]$ given by $\psi(X,g)$.

Let $\mathcal{A}$ be a set of known functions,
$\mathcal{A}=\{f_1,\ldots, f_p, \ \
f_i:L^2([a,b], P) \to \mathbb R \},$ and we denote by

 $$ \textbf{f}(X)=: {\bf{f}}=:(f_1,\ldots, f_p):=\{f_1(X),
\ldots, f_p(X)\},$$
the random vector where each coordinate is the evaluation of the trajectory $X$ on the function $f_j \in \mathcal{A}$. \

Some examples of interest are the following:
\begin{enumerate}
\item \textit{Pointwise evaluation:} $f_1(X) = X(t_1), \ldots, f_p(X) = X(t_p) \mbox{ for } a \leq t_1 < t_2 < \ldots < t_p \leq b.$
\item \textit{Local Averages:} $f_1(X) = \frac{1}{|T_1|}\int_{T_1}X(t)dt, \ldots, f_p(X)= \frac{1}{|T_p|}\int_{T_p}X(t)dt,$ where $\{T_1,
\ldots, T_p \}$ are disjoint intervals of the interval $[a,b]$.
\item \textit{Occupation Measure:} $f_1(X) = \vert \{t: X(t) \in T_1\}\vert , \ldots, f_p(X) = \vert \{t: X(t)
\in T_p\}\vert,$ where $T_1, \ldots, T_p$ are disjoint intervals (not
necessarily bounded) in $\mathbb R$  and $\vert \cdot \vert$ stands
for the Lebesgue measure. That is, the time the
process spends at each interval, or above or below a barrier.
\item \textit{Number of up crossings to a level:}
 Let $f_i(X)= N_i$, where $N_i$ is the number of up crossings of $X$ to a level $c_i \in \mathbb{R}, i=1,\dots,p$.
Recall that the process $X(t)$ is said to have an up crossing of
$c$ at $t_0 >0$ if for some $\epsilon >0$, $X(t) \leq c$  if $t
\in (t_0 - \epsilon, t_0)$ and $X(t) \geq c$ if $t \in (t_0,
t_0+\epsilon)$. If the trajectories are differentiable, then $N_i
= card \{t \in [a,b]: X(t)=c_i, X'(t) > 0 \}$.

\item \textit{Moments of the norm of the process:} $f_1(X) = E(\Vert X \Vert), \ldots, f_p(X) = E(\Vert X \Vert^p).$
\item \textit{Moments of the process:}
$f_1(X) = \int_a^b X(t,\omega)dP(\omega), \ldots f_p(X) = \int_a^b
X^p(t,\omega)dP(\omega).$
\end{enumerate}

\begin{remark}
The choice of the set of functions $\{f_1(X),\ldots,f_p(X)\},$ depends on the nature of the process $X(t)$ and on  features of the statistical procedure under consideration that are relevant to achieve a better understanding of the practical problem. For instance, pointwise evaluation and local averages reveal information towards the domain of $X(t),$ to detect a subset of points or small intervals at the domain, which explain successfully the phenomena. A typical example are spectrometry studies.   
The occupation measure and the number of up crossings to a level, provide information concerning the image of $X(t),$ these characteristics are relevant when measuring the pulse oximetry in preterm infants. Finally, the last two proposals should be used if the aim is to retain global information of $X(t)$,  these features are very important in trading financial problems.
\end{remark}

These sets of functions provide relevant information of a process towards
the statistical procedure that is being conducted. Our aim is to retain a
subset of $\mathcal{A}$, of cardinality  $d<<p$ that contains  ``almost'' all
the relevant information provided by $\mathbf{f}(X)$ to an specific statistical
analysis (classification, regression, etc). To achieve this goal we extend the
ideas introduced by Fraiman et al \cite{FJS08}. Therein, they proposed
to retrieve relevant information from a multivariate model \textit{blinding} unnecessary variables. It is
clear that in the functional data framework a componentwise approach does not make sense, hence the \textit{blinding} procedure
must be redefined.

 Given a subset of indices $I = \{i_1 < \ldots < i_d\} \subset \{1,\ldots, p\}$ with
 $d\leq p$, consider the random vector $\mathbf{f}(I,X):= \mathbf{f}(I):=(f_{i_1}(X),\ldots,f_{i_d}(X)) \in \mathbb R^d.$

\begin{defn} Let $I$ be a subset of $\{1,\ldots, p\}$ denote the \textbf{stochastic blinded process} of $X$, based on $f(I),$ to the process $Z(I):[a,b] \to \mathbb R$ given by
\begin{equation}
\label{blinded}
Z(I)(t)=E(X(t)| \mathbf f(I)):=\eta(t, \mathbf f(I,X)).
\end{equation}
\end{defn}
We denote $Q(I)$ to the distribution of $Z(I).$
\begin{remark}
$Z(I)(t)$ is an stochastic process even though it is the conditional expectation given a random vector $\mathbf f(I) \in \mathbb R^d.$
\end{remark}

Given  a fixed integer $d$, $1\leq d << p$, we let $\mathcal{I}_d$ be
the family of all subsets of $\{1, \ldots, p\}$ with cardinality smaller than or equal to
$d$.

We seek a small subset, $I$, such that $\psi(X,g)$ is as close
as possible to $\psi(Z(I),g)$. The notion of closeness may vary
from one problem to another, and is denoted by $h(I, P,
Q(I), \psi):=h(I)$.

More precisely, $\mathcal{I}_{0} \subset \mathcal{I}_d$ is defined
as the family of subsets in which the minimum $h(I)$ is attained
for $I \in \mathcal{I}_d$, i.e.,
\begin{equation}
\label{objetivo1} \mathcal{I}_{0} = argmin_{I \in \mathcal{I}_d}
h(I).
\end{equation}

In practice, we look for consistent estimates of the set $I_0$, $I_0 \subseteq
\mathcal{I}_0$ based on  a sample $ X_1, \ldots, X_n$ of
iid trajectories of the stochastic process $X$.

Given a subset $I \in \mathcal{I}_d$, the first step is to obtain
the blinded version of the sample, $\hat{X}_1(I), \ldots,
\hat{X}_n(I),$ based on $\mathbf f(I,X)$, using
nonparametric estimates of the conditional expectation.

We denote by $Q_n(I)$ to the empirical distribution of
$\{\hat{X}_j(I) , 1 \leq j \leq n\}$.

For instance, we may consider the $r$ --nearest
neighbor (r-NN) estimates. We fix an integer value $r$, the number
of nearest neighbors used. For each $j \in \{1, \ldots
, n\}$, we find the set of indices $C_j$ of the $r$ nearest
neighbors of $\mathbf f(I, X_j)$ among $\{\mathbf f(I,X_1), \ldots,
\mathbf f(I,X_n)\}$. Next we define the predicted sample processes
as
\begin{eqnarray*}
\hat{X}_j(I)(t)=\hat{E}_P(X_j(t)|\mathbf f(I))=\frac{1}{r}\sum_{m
\in C_j} X_m(t) \in L^2[a,b].
\end{eqnarray*}

Note that the set $C_j$ does not depend on $t$.

Then, given a subset of indices $I \in \mathcal{I}_d$, we define the
empirical version of the objective function $h_n(I)$, and we seek
the optimal subsets of variables $\mathcal{I}_0 \subset \mathcal{I}_d$, which are the
family of subsets in which the minimum of $h_n(I)$ is attained,
i.e.,

\begin{equation}
\label{objetivoempiricofcl} \mathcal{I}_{n} = argmin_{I \in
\mathcal{I}_d} h_n(I).
\end{equation}

This will become clear through the next sections
where we consider different statistical problems and models under
this approach.

In order to prove the consistency of (\ref{objetivoempiricofcl}) to (\ref{objetivo1}) the following hypothesis must be satisfied.

\begin{itemize}
\item[$\mathbf{H1.}$] Let  $\hat{X}(I)$ be a strong consistent estimate of $Z(I)$, i.e.,
 $\left\|Z(I)-\hat{X}(I)\right\|_{L^2[a,b]}\rightarrow_{a.s.}0.$
\end{itemize}

Liam et al \cite{L11} establish general conditions to obtain consistent estimates of the conditional expectation on a space provided with a pseudometric. Hence, \textbf{H1} is a particular case of Liam's et al result, since we are dealing with variables in $\mathbb R^d.$ To be more precise, the conditions imposed by Liam et al can be relaxed, obtaining the following result:

\begin{prop} \label{liamprop}
Let $Z(t) = \eta(t,\mathbf{f}(I,X)) + e$, $Z(t)
\in \mathcal H$, be a separable  Hilbert space,  $\mathbf{f}(I,X)
\in \mathbb R^d$ and $e \in \mathcal H$ a random element with zero
mean and independent of $\mathbf{f}(I,X)$. Let $\eta_n$ be the
non--parametric estimate of the conditional expectation given by
the $r$ nearest neighbors . Given that,
\begin{itemize}
\item[1.] $\mathbf{f}(I,X)$ has density function $g$
such that $0<c_1 \leq g(x) \leq c_2$  for all $x$ in the support of  $\mathbf{f}(I,X)$,
\item[2.] \begin{itemize}
\item[(a)] $\left\Vert \eta(t,\mathbf{f}(I,X)) \right\Vert_{\mathcal H} \leq B \ \ \text{ for all } \mathbf{f}(I,X) \in \mathbb R^d,$
\item[(b)] $\left\Vert \eta(t,\mathbf{f}(I,X)) - \eta(t,\mathbf{f}(I,X)') \right\Vert_{\mathcal H} \leq M \left\Vert \mathbf{f}(I,X) - \mathbf{f}(I,X)' \right\Vert_2$
(Lipschitz condition),
\end{itemize}
\item[3.] there exist $\delta >0$ such that $E\left(\Vert e \Vert_{\mathcal H} ^{2 + \delta}\right)< \infty,$
\item[4.] $k/n \to 0$, $k/log n \to \infty,$
$\sum_{n=1}^{\infty} \left(log n/k\right)^{\delta/2} < \infty,$
\end{itemize}
then
$$
\Vert \eta_n(t,\mathbf{f}(I,X)) - \eta(t,\mathbf{f}(I,X))\Vert_{\mathcal H} = O\left(
\left(\frac{k}{n}\right)^{1/d} + \sqrt{\frac{log n}{k}}\right) \ \
a.s.
$$
\end{prop}

\begin{remark} Proposition \ref{liamprop} is a direct consequence of Theorem 1 from Liam et al \cite{L11} and also from the fact that, in  $\mathbb R^d,$ under condition 1 due to Lebesgue's Differentiation Theorem, we have that $P\left(B(x,h)\right)= O\left(h^d\right)$. In what follows $\mathcal{H}=L^2[a,b].$
\end{remark}

\begin{remark}
 Kudrazow and Vieu \cite{KV13} established general uniform consistency results and convergence rates for $r$-NN generalized regression estimators, where the observed variable takes values in an abstract space. We could have also applied these results to achieve consistency.
\end{remark}

Once we have settled the general framework we proceed to define explicitly the theoretical and empirical objective functions for classification, principal components and regression.


\section{Supervised and Unsupervised Classification}

Let $X(t) \in L^2[a,b]$ be a random process and $K$ the number of groups. For supervised classification, as well as unsupervised
classification (when the number of clusters $K$ is known) we have
a function $g:L^2[a,b]\rightarrow\{1,\ldots,K\}$ that determines
to which group (or cluster) each trajectory belongs to. We denote
the space partition by $G_k=g^{-1}(k), \quad  k=1,\ldots, K$.

For a fixed integer $d<p$, our aim is to find a set $I\subset\{1,\ldots,p\}, \quad \sharp I \leq d,$ where the population objective function, given by,

\begin{eqnarray}
h(I)= 1 - \sum_{k=1}^{K}P(g(X)=k,g(Z(I))=k),
 \label{hclf}
\end{eqnarray}
attains its minimum. Function (\ref{hclf}) measures the difference between the partition of the space considering the original and the blinded trajectory.
\begin{remark}
It is clear that instead of minimizing (\ref{hclf}) one could maximize $h^*(I)= \sum_{k=1}^{K}P(g(X)=k,g(Z(I))=k),$
which is the objective function defined by Fraiman et al \cite{FJS08} for the multivariate case.
\end{remark}

The empirical version of equation (\ref{hclf}) is given by

\begin{eqnarray}
h_n(I) =1- \frac{1}{n} \sum_{k=1}^K \sum_{j=1}^n I_{\{g_n(X_j)=k\}}I_{\{g_n(\hat{X}_j(I))=k\}},
\label{hclempf}
\end{eqnarray}
where the function $g_n:L^2[a,b]\rightarrow\{1,\ldots,K\}$
is the empirical space partitioning function, i.e., it determines to which group (partition subset) each empirical
process belongs. In this case the partition of the space is given by $G_k^{(n)}=g_n^{-1}(k), \quad k=1, \ldots, K$, meaning that (\ref{hclempf}) measures the proportion of observations that are classified in different groups by the original and the blinded processes.

To prove consistency results, in addition to \textbf{H1}, the following assumptions are requested.

\begin{itemize}
\item[$\mathbf{HC1.}$] \begin{itemize}
\item[(a)] The partition of the space is strongly consistent, i.e.,
 given  $\epsilon>0$,  there exists a set
$A(\epsilon)\subset L^2[a,b]$,  with $P(X\in A(\epsilon))>1-\epsilon$  such that, for all $r>0$, $sup_{x \in C(\epsilon,r)}\left|I_{\left\{g_n(x)=k\right\}}-I_{\left\{g(x)=k\right\}}\right|\rightarrow_{a.s.}0$
 as $n\to\infty$ for $k=1,\ldots, K$,
where $C(\epsilon,r)=A(\epsilon)\cap \mathcal K_r$ with $\mathcal K_r$
 is an increasing sequence of compact sets, such that  $P(X \in \mathcal K_r) \to 1$ when $r\to \infty$.
 \item[(b)] $d(X,\partial G_k^n)-d(X,\partial G_k)\rightarrow_{a.s.} 0$ when $n \to \infty$, where $\partial G_k$ (respec. $\partial G_k^n$) is the boundary of $G_k$ (respec. $G_k^n$).
\end{itemize}
\item[$\mathbf{HC2.}$] $P(d(Z(I),\partial G_k) < \delta)\rightarrow 0$
when $\delta\rightarrow 0$ for $k=1, \ldots, K.$
\item[$\mathbf{HC3.}$] The distribution is non--degenerated, i.e. for every
$\delta> 0$,  $P(X \in B(x, \delta))>0$ for almost every $x\in L^2[a,b]$.
\end{itemize}

\begin{theo} \label{consistenciacl} Let $\left\{X_j(t):t \in [a,b] \right\}$ be iid realizations of the stochastic process $X(t)$.
Given $d,$ $1\leq d\leq p$, let $\mathcal{I}_d$ be
the family of all the subsets  $\{1,\dots,p \}$ with cardinality smaller than or equal to $d$ and
let $\mathcal{I}_{0}\subset \mathcal{I}_d$ be the family of all the subsets for which the minimum of equation (\ref{hclf}) is attained.
Under $\mathbf{H1}$, $\mathbf{HC1}$, $\mathbf{HC2}$ and $\mathbf{HC3}$ we have that for each $I_{n} \in \mathcal{I}_{n}$, there exists  $n_0(\omega)$ such that,  for every $n
> n_0(\omega)$, 
with probability one $I_n \in \mathcal{I}_{0}$.
\end{theo}
The proof is given in the Appendix.
\section{Principal Components}
Let $\{X(t):t \in [a,b]\}$ be a stochastic process with finite second moment $\left( E\left(\left\|X^2\right\|_{L^2[a,b]}\right)<\infty \right)$, for almost every $t \in [a,b]$,
with continuous trajectories.  Without loss of generality we assume that it has zero mean $\left( E(X(t))=0\right)$.  We denote
$\nu(t,s)=E(X(t)X(s))$
its covariance function.  As in the finite-dimensional case, the covariance function has associated a linear operator, $\mathbf{\Gamma}:L^2[a,b]\rightarrow L^2[a,b]$ defined as
\begin{eqnarray}
(\mathbf{\Gamma}u)(t)=\int_a^b \nu(t,s)u(s)ds \ \mbox{ for all } u \in L^2[a,b].
\label{defoperador}
\end{eqnarray}
We  assume that
$
\|\nu\|^2_{L^2[a,b]\times L^2[a,b]} = \int_a^b \int_a^b \nu^2(t,s) dt ds < \infty.
$
Cauchy--Schwartz inequality implies that $|\mathbf{\Gamma}u|^2\leq||\nu||^2|u|^2,$
where $|u|$ stands for the standard norm in the space $L^2[a,b],$ while $||\nu||$
denotes the norm in the space $L^2([a,b]\times[a,b])$.  Therefore, $\mathbf{\Gamma}$
is a self-adjoint continuous linear operator. Moreover, $\mathbf{\Gamma}$ is a Hilbert-Schmidt operator.

$\mathcal F$ stands for the Hilbert space of such operators
with inner product given by
$$
\langle\mathbf{\Gamma_1},\mathbf{\Gamma_2}\rangle_{\mathcal{ F}}=trace(\mathbf{\Gamma_1}\mathbf{\Gamma_2})=\sum_{j=1}^\infty \langle \mathbf{\Gamma_1}u_j,\mathbf{\Gamma_2}u_j \rangle_{L^2[a,b]},
$$
where $\{u_j:j\geq1\}$ is any orthonormal basis of $L^2[a,b]$ and
$\langle u,v \rangle$ denotes the ordinary inner product in $L^2[a,b].$

If we consider the basis of the eigenfunctions of  $\mathbf{\Gamma}$, $\{\phi_j:j\geq1\}$, we have that
$
\|\mathbf{\Gamma}\|^2_\mathcal
F=\langle \mathbf{\Gamma},\mathbf{\Gamma} \rangle_{\mathcal F}=\sum_{j=1}^\infty
\lambda_j^2 = \int_a^b \int_a^b \nu^2(t,s)dtds < \infty,
$
where $\{\lambda_j:j\geq1\}$ are the corresponding eigenvalues of
$\mathbf{\Gamma}$. In what follows we assume
that all eigenvalues are different.
For any random variable, $U$, defined as a linear combination of the process $\{X(s)\}$, i.e.
$
U=\int_a^b \alpha(t) X(t) dt= \langle \alpha, X \rangle, \ \alpha \in L^2[a,b],
$
we have that
$
var(U)=E(U^2)=\int_a^b\int_a^b \alpha(t) \nu (t,s) \alpha(s) ds dt= \langle \alpha,\mathbf{\Gamma}\alpha \rangle.
$

The first principal component is defined as the random variable $U_1=\left<\alpha_1,X\right>_{L^2[a,b]}$, such that,
\begin{eqnarray*}
Var(U_1)= \sup_{\|\alpha\|_{L^2[a,b]}=1} Var\left(\left<\alpha,X\right>_{L^2[a,b]}\right)
=\sup_{\|\alpha\|_{L^2[a,b]}=1}\left<\alpha,\mathbf{\Gamma}\alpha\right>_{L^2[a,b]},
\end{eqnarray*}
and the $k-th$ principal component as the variable
\begin{eqnarray}
\label{defpcafun}
U_k=\left<\alpha_k,X\right>_{L^2[a,b]},
\end{eqnarray}
such that \begin{eqnarray*}
Var(U_k)&=& \sup Var\left(\left<\alpha,X\right>_{L^2[a,b]}\right)
=\sup \left<\alpha,\mathbf{\Gamma}\alpha\right>_{L^2[a,b]}
\\&&\text{subject to } \left\|\alpha\right\|_{L^2[a,b]}=1 \text{ and } \left< \alpha,\alpha_j\right>_{L^2[a,b]}=0 \text{ for } j = 1, \ldots, k-1.
\nonumber
\end{eqnarray*}

Therefore, if $\lambda_j > \lambda_{j+1},$ are the eigenvalues of $\Gamma$,  Riesz Theorem \cite{RN65}
entails that the principal components are obtained from the corresponding eigenfunctions. Let
 $\left\{\alpha_k, k\geq1\right\}$ be the basis of eigenfunctions of the covariance linear operator $\mathbf{\Gamma}$, then $U_k=\left<\alpha_k,X\right>$ is the $k-th$ principal component and $Var(U_k)=\lambda_k$.



Assuming that the first $l<p$ principal components yield a good representation of the original process, for each $I \in \mathcal{I}_d,$ we define,
$U_k(I)=\left<\alpha_k,Z(I)\right>_{L^2[a,b]}.$
Our goal is to find a subset $I$ such that $U_k(I)$ is as closest as possible to $U_k,$ for every $k =1,\dots,l$.
The objective function is given by,
\begin{equation}
h(I) = \sum_{k=1}^l  E\left((U_k-U_k(I))^2\right).
\label{hpcafun}
\end{equation}
This function measures the mean value of the square distance between the projection with the original trajectory and with the blinded one.
Given $d<p$,  our aim is to find the set $I \in \mathcal I_d$ that minimizes the objective function (\ref{hpcafun}).

In order to give the empirical version of (\ref{hpcafun}), we shall give  the empirical counterpart of the principal components. Then,
let $\nu_n (t,s)$ be the empirical covariance function, i.e.,
$
\nu_n(t,s)=\frac{1}{n} \sum_{j=1}^n X_j(t)X_j(s),
$
and $\mathbf{\Gamma}_n$ its corresponding linear operator, given by
\begin{eqnarray}
\label{estimadorgamma}
\mathbf{\Gamma}_n = \frac{1}{n} \sum_{j=1}^n \mathbf{V}_j,
\end{eqnarray}
where $\mathbf{V}_j$ is the linear operator defined as
$
\left(\mathbf{V}_j u\right)(t)=\int_a^b X_j(t) X_j(s) u(s) ds.
$
Hence, Fubini's Theorem entails that,  $E\left(\mathbf{V}_j\right)=\mathbf{\Gamma} $ for all $ 1\leq j \leq n.$

Then, the empirical version of function (\ref{hpcafun}) is
\begin{equation}
\label{hnpcafun}
h_n(I)= \sum_{k=1}^l \frac{1}{n}\sum_{j=1}^n \left(U^j_k - U^j_k(I)\right)^2 ,
\end{equation}
where $U^j_k=\left<\alpha_k^n,X_j\right>_{L^2[a,b]}$ and
$U^j_k(I)=\left<\alpha_k^n,\hat{X}_j(I)\right>_{L^2[a,b]}$. The weights of the   $k-th$ empirical principal component for the random sample
 $\left\{X_1,\ldots,X_n\right\}$ are $\alpha_k^n.$  In addition, as a consequence of
 Riezs's Theorem, we have that, $\left\{\alpha_k^n, k\geq 1\right\}$ is the  basis of eigenfunctions  of $\mathbf{\Gamma}_n$ associated to $\left\{\lambda_k^n, k\geq 1\right\}$.

Our aim is to find the set $I\in \mathcal{I}_d$ that  minimizes (\ref{hnpcafun}).

To obtain consistency results, in addition to  \textbf{H1}, the following conditions are needed:
\begin{itemize}
\item[$\mathbf{H2.}$] $E\left(\left\|X-Z(I)\right\|^2_{L^2[a,b]}\right)<\infty$.
\item[$\mathbf{HP1.}$] $E\left(\left\|X\right\|_{L^2[a,b]}^2\right)<\infty$ and
$\left\|\nu\right\|_{L^2([a,b]\times[a,b])} = \int_a^b \int_a^b \nu^2(t,s) dt ds < \infty.$

\end{itemize}

\begin{theo} \label{consistenciapc} Let $\{X_j(t):t \in [a,b] \}$ be a stochastic process
satisfying (\ref{defpcafun}). Given $d, 1\leq d\leq
p$, let $\mathcal{I}_d$ be the family of all the subsets of $\{1,\dots,p \}$
with cardinality smaller than or equal to $d$ and let $\mathcal{I}_{0}\subset \mathcal{I}_d$ be the family of
subsets in which the minimum of equation (\ref{hpcafun}) is obtained.
Then, under $\mathbf{H1,H2}$ and $\mathbf{HP1}$ we have that for each $I_n \in \mathcal{I}_n,$
$I_{n} \in \mathcal{I}_{0}$ ultimately, i.e. $I_{n}=I_0$
with $I_0 \in \mathcal{I}_{0} $ for all $n> n_0(\omega)$, with probability one.
\end{theo}
The proof is given in the Appendix.
\section{The functional linear model}

In the functional linear model the response can be either scalar or functional, in this section we analyze both cases.

\subsection{Linear Model with Scalar Response}

Let $Y\in \mathbb R$ and $X \in L^2[a,b]$, the linear model with scalar response is given by,
\begin{equation}
\label{modelolinealfun}
Y = \int_{a}^b \beta(t)X(t)dt + \varepsilon,
\end{equation}
where $\beta \in L^2[a,b]$ and $\varepsilon$ is a random variable such that $E(\varepsilon)=0$ , $E(\varepsilon^2)< \infty$ and $E(X(t)\varepsilon)=0,$ for almost every $t\in[a,b]$.

Let $\beta_0\in L^2[a,b]$ such that
\begin{equation}
\label{modelolinealfunbeta}
\beta_0 = arg min_{\beta \in L^2[a,b]} E\left(\left(Y- \int_{a}^b \beta(t)X(t)dt\right)^2\right).
\end{equation}
To ensure existence and uniqueness of $\beta_0$ under this setting additional conditions are required.

Assume that
$E\left(\|X\|_{L^2[a,b]}^2\right)<\infty$.
Without loss of generality,  we may assume that the stochastic process is  centered, i.e., $E(X(t))=0$ for almost every
$t \in [a,b]$.  As a consequence of Fubini's Theorem, is clear that $E(Y)=0$, since $E(X(t))=0$ and $E(\varepsilon)=0$.

Let $\mathbf{\Gamma}$ be the covariance operator of the random element $X$ given in (\ref{defoperador}).

Denote $\overline{Im\left(\mathbf{\Gamma}\right)}$ to the closure of $Im\left(\mathbf{\Gamma}\right)=\left\{\mathbf{\Gamma}u, u\in L^2[a,b]\right\}.$ Under mild regular conditions the existence and uniqueness of (\ref{modelolinealfunbeta}) in $\overline{Im\left(\mathbf{\Gamma}\right)}$ is ensured. For a detailed explanation see \cite{FR11}, Chapter 2.

We define the objective function as,
\begin{equation}\label{hlinealfun1}
h(I) = E\left( \left( \int_{a}^b \beta_0(t)X(t)dt - \int_{a}^b
\beta_0(t)Z(I)(t)dt\right)^2 \right).
\end{equation}

It measures the mean square distance between the predicted value considering the original variables and the blinded ones. Given $d<p$, our goal is to find a subset $I\in \mathcal{I}_d$ that minimizes equation  (\ref{hlinealfun1}).

In order to give the empirical counterpart of (\ref{hlinealfun1}) we must provide an estimate of  $\beta.$ There are several ways to face this problem.
One alternative is to estimate the covariance operator $\mathbf{\Gamma}$.
Then project the data onto a finite dimensional space that grows, as the sample size grow, typically by means of the principal components. Sometimes this method is combined with a smoothing procedure.
Another alternative is to represent $\beta$ in a basis of functions of $L^2[a,b]$ satisfying (\ref{modelolinealfunbeta}) adding a penalty to obtain a regular solution. These basis may not be orthonormal, for instance Fourier or Spline basis.

As mentioned in the introduction this problem has been studied by several authors, among them we can mention Cardot et al \cite{CFS03}, Cai and Hall \cite{CH06}, Hall and Horowitz \cite{HH07}, Li and Hsing \cite{LH07}, James et al \cite{JWZ09}.
We consider the estimator proposed by Cardot et al \cite{CFS03}, which is a strong consistent estimate of $\beta$.

The empirical version of the objective function (\ref{hlinealfun1}) is given by
\begin{equation}\label{hnlinealfun1}
h_n(I) = \frac{1}{n}\sum_{j=1}^n \left(\int_{a}^b \beta_n(t)X_j(t)dt -  \int_{a}^b
\beta_n(t)\hat{X}_j(I)(t)dt\right)^2 ,
\end{equation}
where $\beta_n$ is the estimate of $\beta_0$. Our aim is to find the set $I\in \mathcal{I}_d$ that  minimizes
 (\ref{hnlinealfun1}).

\
In order to obtain consistency results, in addition to $\mathbf{H1}$ and $\mathbf{H2}$, the following hypothesis are needed:
\begin{itemize}
\item[$\mathbf{HR1.}$]
$\left\Vert\beta_n-\beta_0\right\Vert_{L^2{[a,b]}}\rightarrow_{a.s.} 0.$
\item[$\mathbf{HR2.}$] $E\left(\Vert X \Vert^2_{L^2[a,b]}\right)< \infty$.
\end{itemize}

It is important to note that the regression estimate, $\beta_n$, introduced in \cite{CFS03}
satisfies $\mathbf{HR1}$.

\begin{theo} \label{consistenciareg1} Let $\left\{\left(Y_j,X_j(t)\right)\in \mathbb{R} \times L^2[a,b] \right\}$ be iid stochastic  processes satisfying  (\ref{modelolinealfun}). Given $d, 1\leq d\leq p$, let $\mathcal{I}_d$ be the family of all the subsets of
$\left\{1,\dots,p \right\}$ with cardinality smaller than or equal to $d$ and let $\mathcal{I}_{0}\subset \mathcal{I}_d$ be the family of al the subsets for which the objective function (\ref{hlinealfun1}) attains its minimum.
Under $\mathbf{H1}$, $\mathbf{H2}$, $\mathbf{HR1}$ and
$\mathbf{HR2}$ we have that for each $I_{n} \in \mathcal{I}_{n}$,
there exists $n_0(\omega)$ such that for each $n > n_0(\omega)$,
with probability one, $I_n \in
\mathcal{I}_{0}$.
\end{theo}

The proof is given in the Appendix.

\subsection{Linear Model with Functional Response}

Let  $Y \in L^2[c,d]$ and $X \in L^2[a,b]$,  the regression model with functional response is given by,
\begin{equation} \label{modelolinealfunfun}
Y(s) = \int_{a}^b \beta(t,s)X(t)dt + \varepsilon(s) \quad s\in[c,d],
\end{equation}
where $\beta \in L^2\left([a,b]\times[c,d]\right)$ and $\varepsilon(s)$ is a random variable for each  $s,$ such that $E\left(\varepsilon(s)\right)=0$ and $E\left(X(t)\varepsilon(s)\right)=0,$ for almost every $t\in[a,b]$, $s\in[c,d]$.

Let $\beta_0\in L^2\left([a,b]\times[c,d]\right)$ such that
\begin{equation*}
\beta_0 = arg \min_{\beta \in L^2\left([a,b]\times[c,d]\right)} E\left(\left\|Y- \int_{a}^b \beta(t,\cdot)X(t)dt\right\|^2_{L^2[c,d]}\right).
\end{equation*}
Under this setting existence and uniqueness of $\beta_0$ are not provided unless some additional hypothesis are established. Following the same ideas as in the case of scalar response these properties can be obtained.

Without loss of generality, we may assume  that $E(X(t))=0$ for almost every $t \in[a,b]$, which entails that $E(Y(s))=0,$ for almost every
$s \in[c,d],$ and  also that $E\left(\|X\|^2_{L^2[a,b]}\right)<\infty$  and
$E\left(\|Y\|^2_{L^2[c,d]}\right)<\infty$.
Let $\mathbf{\Gamma}_X$ be the covariance operator of $X.$

Then under mild conditions,  $\beta_0 \in L^2([a,b]\times[c,d])$ and also the existence and uniqueness of the solution in the orthogonal  space of the  kernel of the covariance operator of $X$ are obtained. See \cite{FR11}, Chapter 2.
%

Once more, following the same ideas as in the case of scalar response, the objective function is given by,
\begin{equation}
\label{hlinealfun2}
h(I) = E\left( \left\| \int_a^b  \beta_0(t,s)X(t)dt - \int_a^b \beta_0(t,s)Z(I)(t)dt \right\|^2_{L^2[c,d]} \right).
\end{equation}

This function measures the mean square distance between the predicted functions considering the original process and the blinded one. Then, given $d<p$, we search for a subset $I\in \mathcal{I}_d$ that minimizes the objective function (\ref{hlinealfun2}).


To give the empirical version of (\ref{hlinealfun2}), we need  an estimator of $\beta$, this problem has been studied by several authors, for instance, Yao et al \cite{YMW05} and M\"{u}ller and Yao \cite{MY08}.
Yao et al, introduce the following estimate
\begin{eqnarray*}
\beta_n(t,s)=\sum_{j'=1}^{J'} \sum_{j=1}^J \frac{\sigma^n_{jj'}\alpha^n_{X,j}(t)\alpha^n_{Y,j'}(s)}{\lambda^n_{X,j}},
\end{eqnarray*}
where $\sigma^n_{jj'}$ is an estimate of $E\left(\xi_{X,j}\xi_{Y,j'}\right)$.
They show that $\beta_n$ converges in probability to $\beta_0$.
Hence, the empirical version of (\ref{hlinealfun2}) is
\begin{equation}\label{hnlinealfunc2}
h_n(I) =\frac{1}{n}\sum_{j=1}^n \left\|\int_a^b  \beta_n(t,s)X_j(t)dt - \int_a^b
\beta_n(t,s)\hat{X}_j(I)(t)dt \right\|^2_{L^2[c,d]}.
\end{equation}

Our goal is to find a subset $I \in \mathcal{I}_d$, that minimizes (\ref{hnlinealfunc2}).
To obtain convergence results, in addition to $\mathbf{H1}$ and $\mathbf{H2}$,  we need the following  conditions:
\begin{itemize}
\item[$\mathbf{HRF1.}$]
$\left\|\beta_n-\beta_0\right\|_{L^2\left([a,b]\times[c,d]\right)}
\rightarrow_{p.} 0.$
\item[$\mathbf{HRF2.}$] $E\left(\|X\|^2_{L^2[a,b]}\right)<\infty$  y
$E\left(\|Y\|^2_{L^2[c,d]}\right)<\infty$.
\end{itemize}
It is important to remark that the estimate $\beta_n$, introduced by Yao et al \cite{YMW05} satisfies $\mathbf{HRF1.}$

Then we have the following consistency result.
\begin{theo} \label{consistencialineal2fun} Let $\left\{\left(Y_j(s),X_j(t)\right)\in L^2[c,d] \times L^2[a,b] \right\}$ be iid stochastic processes
satisfying (\ref{modelolinealfunfun}). Given $d, 1\leq d\leq p$, let $\mathcal{I}_d$ be the family of all the subsets of $\left\{1,\dots,p \right\}$ with cardinality smaller than or equal to $d$ and let
$\mathcal{I}_{0}\subset \mathcal{I}_d$ be the family of all the subsets where the minimum of the function (\ref{hlinealfun2}) is attained. Under $\mathbf{H1}$,
$\mathbf{H2}$, $\mathbf{HRF1}$ and $\mathbf{HRF2}$ given $I_n \in \mathcal{I}_n$ it can be shown that
$P\left(I_n \in \mathcal I_0\right) \to 1$. 
\end{theo}

The proof is given in the Appendix.
\begin{remark}
If $\beta_n$ is a strong consistent estimate of $\beta_0$, then replacing hypothesis $\mathbf{HRF1}$ by
\begin{itemize}
\item[$\mathbf{HRF1^*.}$] $\left\|\beta_n-\beta_0\right\|_{L^2\left([a,b]\times[c,d]\right)} \rightarrow_{a.s.} 0$
\end{itemize}
strong consistency is  attained in Theorem \ref{consistencialineal2fun}.
\end{remark}
\section{Numerical Aspects}

This section has two main parts. First we introduce a stochastic algorithm to find the optimal subset of functions. Then we illustrate the performance of our proposal analyzing some well known data sets.

\subsection{The algorithm}

As mentioned in Section \ref{main} our aim is to select, $I_d$, a subset of $\mathcal{A}=\{f_1,\dots,f_p\},$ that better captures the information of a stochastic process relative to a statistical procedure, i.e., $h_n(I_d)$ must be ``small''.

Before describing the algorithm we shall discuss when to consider that $h_n(I_d)$ is small enough.
For the case of classification since $h_n(I_d)$ is the proportion of observations classified into different groups when we use the original or the blinded trajectories, i.e. the ``matching error rate", it is clear how to establish that $h_n(I_d)$ is small enough. However for the cases of regression and PCA $h_n(I_d)$ depends on the units
 in which the functions are measured (regression and PCA). Therefore,  our proposal is to rescale the objective function, $\widetilde{h}_n$, to make it independent of the units of the data.
Then, a subset of functions will be chosen if
\begin{equation}
\widetilde{h}_n(I_d)< \epsilon,
\label{hnrescalada}
\end{equation}
where $\epsilon$ is a positive constant that must be supplied by the user.

We exhibit the rescaled objective function for each of the problems analyzed in this work.
%

For the case of principal components we suggest to rescale the objective function as follows,
$$
\widetilde{h}_n(I_d)=\frac{h_n(I_d)}{\sum_{k=1}^{l}\frac{1}{n}\sum_{j=1}^{n}\left(U_k^j\right)^2},
$$
where $h_n(I_d)$ is given by equation (\ref{hnpcafun}).
In an analogue way, the rescaled objective function for the linear functional model with scalar response is given by,
$$
\widetilde{h}_n(I_d)=\frac{h_n(I_d)}{\frac{1}{n}\sum_{j=1}^n \left(\int_a^b \beta_n(t)X_j(t)  dt\right)^2},
$$
where $h_n(I_d)$ is given by equation (\ref{hnlinealfun1}).
Finally, we consider the linear regression problem when the response is functional the rescaled objective function is,

$$
\widetilde{h}_n(I_d)=\frac{h_n(I_d)}{\frac{1}{n}\sum_{j=1}^n \left\|\int_a^b \beta_n(t,s)X_j(t)dt \right\|^2_{L^2[c,d]}},
$$
where $h_n(I_d)$ is defined by equation (\ref{hnlinealfunc2}). In the last three cases the rescaled objective functions measure  the relative difference between the procedure carried out with the blinded processes and with the original processes.

\begin{remark}
For classification $\widetilde{h}_n(I_d)\equiv h_n(I_d),$ and $\epsilon$ is the maximum  matching error rate allowed by the user.
\end{remark}

  It is clear that if $p$ is moderate or large it is not feasible to analyze the $2^p-1$ subsets of functions, hence we need to provide a numerical strategy to tackle this problem.

We introduce a forward algorithm that has two main steps. On the first one, an exhaustive search over all the subsets with cardinality $d_1$ (where $d_1<<d$) is performed. While the second step is a forward stochastic search.

\textbf{Exhaustive Step}

\hfill\begin{minipage}{\dimexpr\textwidth-1cm}
Consider all the subsets of functions with cardinality smaller than or equal to $d_1$. Keep the $N_0$ subsets with  smaller $\widetilde{h}_n(\widetilde{I}_{\widetilde{d}}),$ where $\widetilde{I}_{\widetilde{d}}=\widetilde{I}_1,\dots,\widetilde{I}_{N_0}.$
\end{minipage}
\hfill\begin{minipage}{\dimexpr\textwidth-2cm}
If there is at least one of them satisfying condition (\ref{hnrescalada}), stop the algorithm.
\end{minipage}

 \hfill\begin{minipage}{\dimexpr\textwidth-3cm}
 Among all the subset satisfying condition (\ref{hnrescalada}) retain the subset with smallest cardinality, in case of ties, keep the subset with smallest $\widetilde{h}_n$.
\end{minipage}
 \hfill\begin{minipage}{\dimexpr\textwidth-2cm}
 Otherwise, if condition (\ref{hnrescalada}) is not satisfied, proceed with the Stochastic Step, where the input are the subsets of functions given by $\widetilde{I}_1,\dots,\widetilde{I}_{N_0}$.
\end{minipage}

\textbf{Stochastic Step}

Repeat until condition (\ref{hnrescalada}) is attained.

\hfill\begin{minipage}{\dimexpr\textwidth-1cm}
 For each subset $\widetilde{I}_{\widetilde{d}},$ where ${\widetilde{d}}=1,\dots,N_0.$
\end{minipage}

\hfill\begin{minipage}{\dimexpr\textwidth-2cm}
 Choose at random without replacement $N_1$, $i_1,\dots,i_{N_1}$ functions from $I \backslash \widetilde{I}_{\widetilde{d}}.$
\end{minipage}

\hfill\begin{minipage}{\dimexpr\textwidth-2cm}
 Consider the subsets $\widetilde{I}_{\widetilde{d},j}=:\widetilde{I}_{\widetilde{d}} \bigcup \{i_j\}$ for $j=1,\dots,N_1.$
\end{minipage}

\hfill\begin{minipage}{\dimexpr\textwidth-3cm}
 Compute $\widetilde{h}_n(\widetilde{I}_{\widetilde{d},j})$    for $j=1,\dots,N_1.$
\end{minipage}

 \hfill\begin{minipage}{\dimexpr\textwidth-3cm}
 Retain the $N_0,$  subsets with smaller  $\widetilde{h}_n$, with a slight abuse of notation denote them  $\widetilde{I}_1,\dots,\widetilde{I}_{N_0}.$
\end{minipage}

\hfill\begin{minipage}{\dimexpr\textwidth-3cm}
If there is at least one of them satisfying condition (\ref{hnrescalada}), stop the algorithm. Among all the subset satisfying condition (\ref{hnrescalada}) retain the subset with smallest cardinality, in case of ties, keep the subset with smallest $\widetilde{h}_n$.
\end{minipage}

\hfill\begin{minipage}{\dimexpr\textwidth-3cm}
 Otherwise, if condition (\ref{hnrescalada}) is not satisfied proceed with a revision step.
\end{minipage}

\hfill\begin{minipage}{\dimexpr\textwidth-4cm}
For each subset $\widetilde{I}_{\widetilde{d}},$ where ${\widetilde{d}}=1,\dots,N_0.$
\end{minipage}

\hfill\begin{minipage}{\dimexpr\textwidth-4cm}
Replace at random, one at a time, each element of $\widetilde{I}_{\widetilde{d}},$ if $\widetilde{h}_n$ decreases keep the best subset.
\end{minipage}

\hfill\begin{minipage}{\dimexpr\textwidth-5cm}
If condition (\ref{hnrescalada}) is attained stop the algorithm.
\end{minipage}

\hfill\begin{minipage}{\dimexpr\textwidth-5cm}
Otherwise, run the Stochastic Step from the top.
\end{minipage}

The algorithm depends on several parameters, namely, $\epsilon, d_1, N_0$ and $ N_1.$ We have already discussed the roll that $\epsilon$ plays. The other three parameters can be easily settled and are all in relation with the cardinality $\mathcal{A.}$ $d_1$ denotes  is the maximum cardinality analyzed in the exhaustive step. This number should be small, specially if the  cardinality of $\mathcal{A}$ is big. $N_0$ is the number of subsets retained after the exhaustive step and $N_1$ is the number of variables added to each potential subset chosen either in the exhaustive step or once the stochastic step has been run without achieving condition (\ref{hnrescalada}), this constants can be moderate, allowing several bifurcations we seek to avoid local minima.

\subsection{Real Data Example}

In this Section we will analyze the behavior of our procedure on several real data examples. Routines are available upon request to the authors.

\textbf{The growth data set}

\begin{figure}
\begin{center}
\includegraphics[width=12cm]{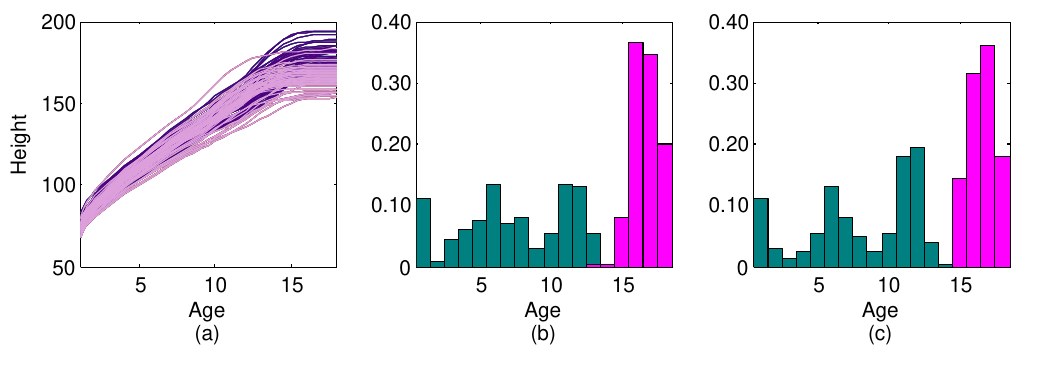}
\end{center}
\vspace{-15pt} \caption{\textbf{(a)} Height curves, the darker ones are from the boys and the lighter ones for the girls. \textbf{(b)} and \textbf{(c)} The histograms show the ages chosen by the blinding procedure, the green ones represent the first time and the purple one the second one, for $r=3$ or $r=5.$}
 \label{growthhistograma}
\end{figure}

The growth data set has been analyzed in several opportunities, and corresponds to  the Berkeley growth study.  It is available in the functional data analysis packages for Matlab and R. The data set consists on the height measurements of 54 girls and 39 boys, that were measured on 31 opportunities from 1 to 18 years. The original data was smoothed by monotonic cubic regression spline, the curves are shown in Figure \ref{growthhistograma} (a).

 Our aim is to classify the growth curves for a sample of boys and girls. Since there is no learning and test sample we shall proceed following the ideas from Lopez-Pintado and Romo \cite{LR06}. Therein, they suggest to separate the data into four groups of the same size and consider one of them as test group. The remaining observations constitute the learning set. The data set is classified  by fitting a functional generalized additive model (see \cite{FO12}) and the number of misclassified observation is computed from the test group. All these steps are repeated changing the test group 4 times and
the total error rate is defined as the number of misclassified observations over the
total number of observations.
It is clear that different partitions yield different misclassification rates, to avoid this effect we repeat this procedure 50 times.
The misclassification rate is between $3.23\%$ and $10.75\%$, the mean is $6.17\%$ and the median is $6.45\%,$ which are very good results in comparison to those obtained by Lopez-Pintado and Romo \cite{LR06}.

Our goal is to find which are the key time instants  that determine that boys and girls have different growth patterns. Then, if the set $\mathcal{A}$ is conformed by the evaluations on the grid of time points, we apply the blinding procedure to these functions. The cardinality of $\mathcal{A}$ is 31, hence it is feasible to perform an exhaustive search. The nonparametric estimation is done by r-nearest neighbors, with $r=3$ or $5$.
If $d_1=1$, the mean matching error is high $21.94\%$ (respc. $22.09\%$) for $r=3$ (respc. $r=5$.) Then we decided to set $d_1=2,$ this means to look for the pair of functions with minimum matching error that belongs to $\mathcal{A}.$ In  these cases the mean matching error decreases to $4.75\%$ (respc. $4.77\%$) for $r=3$ (respc. $r=5$.)
In Figure \ref{growthhistograma} (b) (respc. (c)) we show the histogram for the instants chosen by the blinding procedure, indicating with different colors the first and second choice, in almost every case the final height is a relevant feature, boys tend to be taller than girls. For the other measure there are clearly three modes, the first one is the initial measurement (people tend to grow in the same percentile curve throughout their lives), the second one is close to 6 years (there is an acceleration in the growth speed) and the third one during the puberty (girls grow earlier than boys). Hence the time points chosen by the algorithm reflect important patterns on the growth charts.


It is important to remark that histograms corresponding to  $r=3$ or $r=5$ are very similar.

\textbf{The phoneme data set}

In the following example, the data corresponds to the discretized log-periodograms, which is a widely used method for casting speech data in a form suitable for speech recognition.  The learning data set, as well as the test data sets, contain 250 recordings of men
pronouncing five phonemes. Each curve is recorded on a grid of 150 equally spaced time points. The phonemes are ``sh" as in ``she'', ``iy''  as the vowel in ``she'', ``dcl''  as in ``dark'', ``aa''  as the vowel in ``dark'', and
``ao'' as the first vowel in ``water''. The label of each observation is known.

The dataset is in the R package  \textit{fda.usc} (Functional Data Analysis and Utilities for Statistical Computing) and it is a part of the original one which can be found at \url{http://www-stat.stanford.edu/ElemStatLearn}.

%


The data set is classified  fitting a functional generalized kernel additive models (see \cite{FO12})  and only $6.4\%$ observations of the test sample are misclassified.

The set of functions $\mathcal{A},$ are the evaluations on the 150 grid of time--points.

In order to apply the blinding procedure we must fix the values of the parameters involved at each stage in the algorithm. The maximum matching error rate is  $\epsilon=0.10.$ For the exhaustive step we set $d_1=1,$ since $\mathcal{A}$ has 150 functions. For the stochastic step we used $N_0=3$ or $6$ and $N_1=5$ or $10.$
In addition, for the number of nearest neighbors for the nonparametric estimation we considered either $r=5$ or $10.$
We performed 100 replicates for each parameter configuration.

In Table \ref{phonememedia} we can observe that the mean number of functions chosen is always between 4.5 and 5, in addition we may add that in every case between 3 and 6 functions where selected and that the median number of functions is always either four or five. Moreover, in Table \ref{phonemeerrores} we report the mean matching error rate, which is in every case between $8.2\%$ and $8.5\%$.  The minimum matching error rate in every case is between 4 or 5 $\%.$ The results obtained by the blinding procedure using different values for the parameters are practically the same.

 \begin{table}[h!]
\centering { \begin{tabular}{cccccc}
\hline
& $r$ &  \multicolumn{2}{c}{5} & \multicolumn{2}{c}{10}\\
 \cline{2-6}
 $N_1$ & $N_0$ & $3$ &  $6$ &  $3$ &  $6$ \\
\cline{1-6}
 $5$ &  &  4.93   &  4.64    &  5.02   & 4.67   \\
$10$ &  & 4.68   &  4.47    &  4.79  &  4.47  \\
\hline
\end{tabular}}
\caption{Mean number of functions chosen for each parameter configuration.}
\label{phonememedia}
\end{table}

 \begin{table}[h!]
\centering { \begin{tabular}{cccccc}
\hline
& $r$ &  \multicolumn{2}{c}{5} & \multicolumn{2}{c}{10}\\
 \cline{2-6}
$N_1$ & $N_0$ & $3$ &  $6$ &  $3$ &  $6$ \\
\cline{1-6}
 $5$ &  &  8.48   &  8.31    &  8.44   & 8.34 \\
$10$ &  &  8.32   &  8.20    &  8.34  &  8.36  \\
\hline
\end{tabular}}
\caption{Mean matching error rate.}
\label{phonemeerrores}
\end{table}

\textbf{The Canadian weather data}

This data set has been introduced first by Ramsay and Silverman \cite{RS02} and it is available in the R package \textit{fda} (see \cite{RWGH12}). The data contains daily temperature and rainfall observations over the course of a year measured on 35 monitoring stations from Canada. Several authors analyzed this data set in the context of scalar regression (see \cite{GS14} and references therein), their aim is to predict the log annual precipitation from the temperature observations.

We estimate the functional linear model  as proposed by Cardot et al \cite{CFS03}, i.e. considering  an almost sure consistent estimate. This procedure has been  numerical implemented by Goldsmith and Scheip \cite{GS14}, they show that it  has good empirical performance.

Our aim is to find the time periods where the temperature has greater influence in the prediction of the log annual precipitation amount.
Hence, the set of functions $\mathcal{A},$ that we analyze is conformed by 40 local averages of the temperatures from non--overlapping intervals of 9 days,  $f_1,\dots,f_{40},$ and one local average with the remaining last  five days, $f_{41}$. This means that $f_1$ is the mean temperature from day 1 to 9, $f_2$ is the mean temperature from day 10 to 18 and so on and so forth. We apply the blinding procedure to $\mathcal{A}.$ The cardinality of $\mathcal{A}$ is 41, then it is feasible to perform an exhaustive search. The nonparametric estimation is done by r-nearest neighbors, with $r=3$ or $5$.

\begin{figure}[h!]
\begin{center}
\includegraphics[width=12cm]{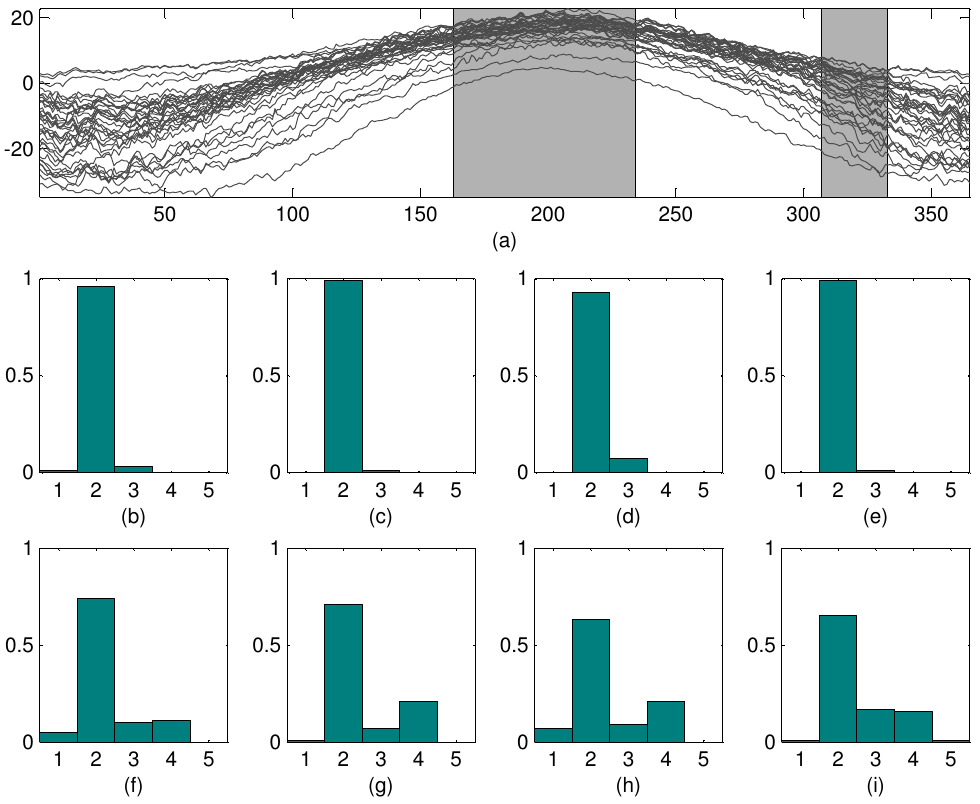}
\end{center}
\vspace{-15pt} \caption{\textbf{(a)} Temperature dataset. \textbf{(b)--(i)}  The histograms show the number of functions chosen by the algorithm for the different parameter choices. \textbf{(b)} $r=3$, $N_0=3,$  $N_1=5$ \textbf{(c)} $k=3$, $N_0=3,$  $N_1=10$ \textbf{(d)} $r=3$, $N_0=6,$  $N_1=5$ \textbf{(e)} $r=3$, $N_0=6,$  $N_1=10$ \textbf{(f)} $r=5,$ $N_0=3,$  $N_1=5$ \textbf{(g)} $r=5$, $N_0=3,$  $N_1=10$ \textbf{(h)} $r=5$, $N_0=6,$  $N_1=5,$ \textbf{(i)} $r=5$, $N_0=6,$  $N_1=10.$}
 \label{histogramacuantosweather}
\end{figure}

In Table \ref{weatherex} we exhibit the optimal functions chosen and also the value of the rescaled objective function, $\widetilde{h},$ for $d=1,2,3$ or $4.$  We observe that $\widetilde{h}$ decreases on $d,$ there is a bigger gain if we choose two functions instead of one, but practically no gain if more functions are chosen. The functions mainly retain information that correspond with two larger periods of time, the first one during the summer and the other one during the autumn. In Figure \ref{histogramacuantosweather} (a), the temperature functions are exhibit, the gray rectangles highlight the areas chosen by the exhaustive procedure.
Even though, there are not much differences for $r=3$ or $5$ nearest neighbors the objective function shows better behavior for $r=3.$

\begin{table}[h!]
\centering { \begin{tabular}{ccccc}
\hline  &  \multicolumn{4}{c}{$r$} \\
&   \multicolumn{2}{c}{3} & \multicolumn{2}{c}{5}\\
 \cline{2-5}
 $d_1$ & $f_i$ & $\widetilde{h}$ & $f_i$ & $\widetilde{h}$ \\
\hline
 $1$   &  35   &  0.0013    &  37   & 0.0013 \\
$2$   &  26, 35   &  0.0010    &  20, 37  &  0.00117  \\
$3$   &  26, 36, 37   &  0.00094    & 19, 35, 37  &  0.00116  \\
$4$   &  24, 35, 37, 38   &  0.00092    & 22, 24, 36, 37  &  0.00113  \\
\hline
\end{tabular}}
\caption{Functions chosen by an exhaustive search with their corresponding $\widetilde{h}$ values.}
\label{weatherex}
\end{table}

In addition, we also run the stochastic algorithm, first we fix the values of the parameters involved at each stage of the algorithm. For the exhaustive step, we set $d_1=1.$ For the stochastic search, we fix, $\epsilon=0.0012,$  $N_0=3$ or $6$ and $N_1=5$ or $10.$ The number of nearest neighbors remains the same as in the exhaustive search.
We perform 100 replicates for each parameter configuration.
For $r=3$ in almost every case two functions are chosen (Figure \ref{histogramacuantosweather} (b) to (e)), while for $r=5,$ even though  the median number of functions is always 2,  there is more dispersion (Figure \ref{histogramacuantosweather} (f) to (i)).

\begin{figure}[h!]
\begin{center}
\includegraphics[width=12cm]{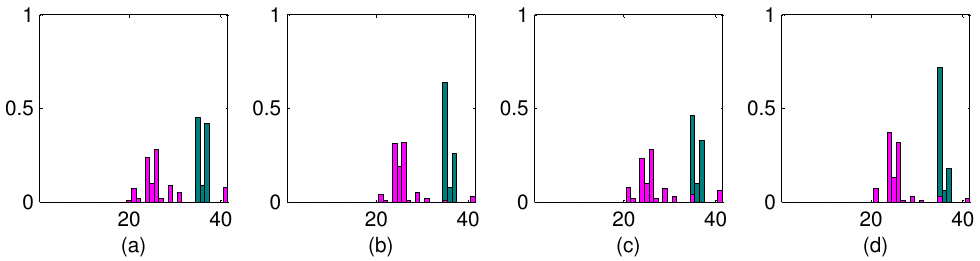}
\end{center}
\vspace{-15pt} \caption{ The histograms show the distribution of functions chosen by the algorithm, for $r=3,$ when two functions are chosen for the different parameter configurations. \textbf{(a)}  $N_0=3,$  $N_1=5$ \textbf{(b)} $N_0=3,$  $N_1=10$ \textbf{(c)}  $N_0=6,$  $N_1=5$ \textbf{(d)}  $N_0=6,$  $N_1=10$ }
 \label{weatherperiods}
\end{figure}

In Figure \ref{weatherperiods} (a) to (d)  we show the histograms for the functions chosen, for $r=3,$ by the algorithm for the different parameter configurations, when two functions are chosen, we indicate with different colors each choice. In every case, one of the periods chosen is during the autumn and in most of the cases the other one is during the summer, this results are similar as those obtained by the exhaustive search. If more than two functions are chosen, in almost every case, two of them lay during the autumn and the remaining during the summer.  The results for $r=5$ are similar.

In addition, we change the set of functions considered in this example, we use the local averages during the twelve months, $\widetilde{\mathcal{A}}$. We run the exhaustive procedure, if $d_1=1$ the month chosen is November, which includes functions $f_{35},f_{36}$ and $f_{37}$ from $\mathcal{A}$. If $d_1=2,$ November is one of the months chosen while the other one is August for $r=3$ and July for  $r=5,$  these months are contained in the subsets of functions $f_{21},\dots,f_{27}$ from $\mathcal{A}$. In every case, the difference between the values of the $\widetilde{h}$ with $\mathcal{A}$ or with $\widetilde{\mathcal{A}}$ is smaller than $0.0002.$ These results are consistent with those obtained by Lee et al \cite{LBP12} and by James et al \cite{JWZ09} that also analyzed this problem in the context of sparse regression.

\begin{figure}[h!]
\begin{center}
\includegraphics[width=12cm]{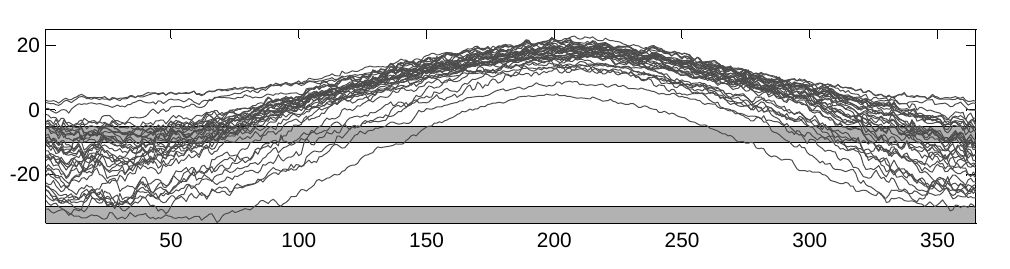}
\end{center}
\vspace{-15pt} \caption{ Temperature bands chosen for $r=3.$ }
 \label{weatherbandasfig}
\end{figure}

Finally, we analyze this data set from a different perspective. Our goal is to determine temperature bands that have greater impact on the prediction of the logarithm of the annual precipitation. Hence, we consider the set of function $\mathcal{A}=\{f_{1},\dots,f_{12}\},$ where $f_{i}$ denotes the occupation measure in a band of $5$ Celcius degrees, i.e., $f_{i}=|\{t: (-35+5(i-1))\leq X(t) < (-35+5i)\}|,$ for $i=1,\dots,12.$ Since the cardinality of $\mathcal{A}$ is small we perform an exhaustive search, once more the nonparametric estimation has been performed for either  $r=3$ or $r=5$ neighbors. In Table \ref{weatherbandas} we exhibit the optimal values of the rescaled objective function,  $\widetilde{h}.$ For $r=5$ it is clear that the maximum gain is achieve if two functions are kept, for $r=3$ it is not clear if two or three functions should be retained. In every case, one of the functions retained correspond to an intermediate temperature, while the other one corresponds to a an extreme value that can be either cold or hot, Figure \ref{weatherbandasfig} is exhibit as example.

\begin{table}[h!]
\centering { \begin{tabular}{ccc}
\hline  &  \multicolumn{2}{c}{$r$} \\ \cline{2-3}
 $d_1$ & 3 & 5 \\
\hline
 $1$   &    0.0036      & 0.0044 \\
$2$   &    0.0030      &  0.0030  \\
$3$   &   0.0027      &  0.0029  \\
$4$   &    0.0025    &  0.0028  \\
\hline
\end{tabular}}
\caption{Optimal values of $\widetilde{h}$ corresponding to the exhaustive search.}
\label{weatherbandas}
\end{table}

\section{Concluding Remark}

In this paper we address the problem of feature selection when the data is functional, we study several
statistical procedures including classification, regression and principal components. One advantage
of the blinding procedure is that it is very flexible since the features are defined by a set of functions,
relevant to the problem being studied, proposed by the user. Our method is consistent under a set of quite general assumptions, and produces good results with the real data examples that we analyze. In any of the statistical procedures studies sparsity has been assumed.

From the numerical aspects, it is clear that if the cardinality of $\mathcal{A}$ is moderate
or large an exhaustive search in unfeasible, hence an stochastic heuristic algorithm is introduced.
The algorithm depends on several parameters, and an optimal way to choose them
is beyond the scope of this paper.

A possible shortcut of our approach is that we
are assuming that the cardinal $p$ of the set of functions $\{f_1,
\ldots, f_p\}$ is fixed, although it may be large. An extension to
the case when $p \to \infty$ - besides a much more involved theory
- would need in particular, some asymptotic results which ensure
that assumption {\bf{H1}} holds. Then, the regression function
should fulfill Besicovitch condition, see for instance Forzani et
al \cite{FFL12}. An important issue is to extend our  results on
feature selection to this framework in the next future.

Several authors, among them  Cuevas \cite{C14} and Aneiros et al \cite{AFV15} suggest
that the link between variable selection methods for high dimensional data and functional data should
be deeply studied. In particular they focus on the idea considering a discrete representation of
the functions and determine whether all the points are relevant for the statistical procedure carried
out or if it is enough to observe some key points. Ferraty et al \cite{FHV10} and Kneip and
Sarda \cite{KS11} consider different approaches to tackle this problem for the regression problem.
We consider that the question the pose is very relevant (and even though in this work we deal with it),
it should be deeply studied. More general statistical problems should be considered and theoretical
 results concerning the number of points that should be chosen should be studied.

\section{Proofs}
\begin{proof} \textbf{(Theorem \ref{consistenciacl})}
Since $\mathcal{I}_d$ is finite it is enough to show that
$
\lim_{n\rightarrow\infty} h_n(I)=h(I) \mbox{ a.s., for all } I \in \mathcal{I}_d.
$
That is,  for  $k=1, \ldots, K$,
\begin{eqnarray}
\lim_{n\to\infty} \frac{1}{n} \sum_{j=1}^n
I_{\{g_n(X_j)=k\}}I_{\{g_n(\hat{X}_j(I))=k\}}=P(g(X)=k,g(Z(I))=k)
\mbox{ a.s.}  \label{convenunk}
\end{eqnarray}

We define the unobservable trajectories $Z_1(I), \ldots, Z_n(I)$, where
\begin{eqnarray}
Z_j (I)(t) = E_P(X_j(t) |\mathbf f(I,X_j)),
\end{eqnarray}
and denote by $Q^*_n(I)$ it's empirical
distribution.

Equation (\ref{convenunk}) holds,  if for every fixed $k$,
\begin{eqnarray}
\lim_{n\rightarrow\infty} \frac{1}{n} \sum_{j=1}^n
I_{\{g_n(X_j)=k\}}I_{\{g_n(Z_j(I))=k\}}=P(g(X)=k,g(Z(I))=k) \mbox{
a.s.} \label{convenunka}
\end{eqnarray}
and
\begin{eqnarray}
\lim_{n\rightarrow\infty} \frac{1}{n} \sum_{j=1}^n
I_{\{g_n(X_j)=k\}}\left[I_{\{g_n(\hat{X}_j(I))=k\}}-I_{\{g_n(Z_j(I))=k\}}\right]=0
\mbox{ a.s..} \label{convenunkb}
\end{eqnarray}

We derive (\ref{convenunka}) from the  following statements,
\begin{eqnarray}
\lim_{n\rightarrow\infty} \frac{1}{n} \sum_{j=1}^n
I_{\{g_n(X_j)=k\}}I_{\{g_n(Z_j(I))=k\}}-I_{\{g(X_j)=k\}}I_{\{g(Z_j(I))=k\}}=0
\mbox{ a.s. } \label{convenunkai}
\end{eqnarray}
and
\begin{eqnarray}
\lim_{n\rightarrow\infty} \frac{1}{n} \sum_{j=1}^n
I_{\{g(X_j)=k\}}I_{\{g(Z_j(I))=k\}}=P(g(X)=k,g(Z(I))=k) \mbox{
a.s.} \label{convenunkaii}
\end{eqnarray}

Equation (\ref{convenunkaii}) follows from de Law of the Large Numbers. To proof (\ref{convenunkai}) we observe that the left hand side of the equation is
dominated by
\begin{eqnarray*}
&&\frac{1}{n} \sum_{\{X_j \in C(\epsilon,r)\}\cap\{Z(I)_j \in
C(\epsilon,r)\}}
|I_{\{g_n(X_j)=k\}}I_{\{g_n(Z_j(I))=k\}}-I_{\{g(X_j)=k\}}I_{\{g(Z_j(I))=k\}}|
+
\\ && \frac{1}{n} \sum_{\{X_j \notin C(\epsilon,r)\}\cup\{Z(I)_j \notin C(\epsilon,r)\}} |I_{\{g_n(X_j)=k\}}
I_{\{g_n(Z_j(I))=k\}}-I_{\{g(X_j)=k\}}I_{\{g(Z_j(I))=k\}}|,
\end{eqnarray*}
where $C(\epsilon,r)$ is given in assumption
\textbf{HC1(a)}.
From \textbf{HC1(a)} it can be seen that the first term  is clear that it converges to zero for every $\epsilon$ and $r$. On the other hand, the last term vanishes due to the Law of Large Numbers.
Then (\ref{convenunkai}) holds, and the proof of  (\ref{convenunka}) is completed.

To proof equation (\ref{convenunkb}), it is enough to show that
\begin{eqnarray}
\sharp \{j: g_n(\hat{X}_j(I))=k, g_n(Z_j(I))\neq k, g_n(X_j)=k\}/n
\rightarrow 0 \ \mbox{a.s.}, \label{convenunkbi}
\end{eqnarray}
and
\begin{eqnarray}
\sharp \{j: g_n(\hat{X}_j(I))\neq k, g_n(Z_j(I))= k,
g_n(X_j)=k\}/n \rightarrow 0 \ \mbox{a.s.} \label{convenunkbii}
\end{eqnarray}

We define the sets B, C and D as follows:

$$B=\{\omega \in \Omega: \|Z_j(I)-\hat{X}_j(I)\|^2_{L^2[0,1]}\rightarrow_{n\rightarrow\infty}0 \ \forall j \},$$
$$C_j=\{\omega \in \Omega: d(X_j,\partial G_k^n)-d(X_j,\partial G_k)\rightarrow 0\} \mbox{ and } C=\cap_{j=1}^\infty C_j,$$
$$D=\{\omega \in \Omega:d(\partial G_k^n,\partial G_k)\rightarrow_{n\rightarrow\infty} 0\}.$$
From \textbf{HC1(b)} and \textbf{H1} we have that $P(B\cap C)=1,$ and from \textbf{HC3} and \textbf{HC1(b)} we have that $P(D)=1.$
Hence, $P(B\cap C \cap D)=1.$ Then, given $\delta>0$
and $\omega \in B \cap C \cap D,$ there exists $n_0=n_0(\omega, \delta)$
such that $max_{j=1, \ldots,n}\|Z_j(I)-\hat{X}_j(I)\|^2_{L^2[0,1]} \leq \delta/2.$

Given $\omega \in B \cap C \cap D,$ $\delta>0$ and $n\geq n_0(\omega, \delta)$ we have that:
\begin{eqnarray*}
\{j:g_n(\hat{X}_j(I))=k,g_n(Z_j(I))\neq k, g_n(X_j)=k\} &\subseteq&  \{j:d(Z_j,\partial G_k)<2\delta\}.
\end{eqnarray*}

%
%

This implies that the left hand side of (\ref{convenunkbi}) is
dominated by
$
\frac{1}{n} \sharp \{j:d(Z_j,\partial G_k)<2\delta\} \leq \frac{1}{n} \sum_{j=1}^n I_{\{d(Z_j,\partial G_k)<2\delta\}},
$ which converges a.s. to $P(d(Z_j,\partial G_k)<2\delta)$ when  $n\rightarrow \infty.$

Finally, from \textbf{HC2} we get that $lim_{\delta \rightarrow 0} P(d(Z,\partial G_k)<2\delta)=0,$ which concludes the proof of (\ref{convenunkbi}). The proof of (\ref{convenunkbii}) is completely analogue.

\end{proof}

\begin{proof}(\textbf{Theorem \ref{consistenciapc}})

In order to prove our statement it is enough to
show that (\ref{hnpcafun}) converges to (\ref{hpcafun}) a.s. To
simplify notation, without loss of generality, we consider only
one principal component (i.e., l=1), which will be denoted by
$\alpha=\alpha _1$. Then we have,
$$
h_n(I) =\frac{1}{n}\sum_{j=1}^n(U^j_1 - U^j_1(I))^2 =\frac{1}{n}\sum_{j=1}^n ((\alpha^n,X_j)-(\alpha^n,\hat{X}_j(I)))^2
=\frac{1}{n}\sum_{j=1}^n\left(\int_a^b\alpha^n(t)(X_j(t)-\hat{X}_j(I)(t))dt\right)^2 ,
$$ and
$$
h(I) = E_P\left( (U_1 - U_1(I) )^2\right)=E_P\left(((\alpha, X)-(\alpha,Z(I)))^2\right)= E_P\left(\left(\int_a^b\alpha(t)(X(t)-Z(I)(t))dt \right)^2\right).
$$
We define the unobservable functions $Z_1(I), \ldots, Z_n(I)$, where
\begin{eqnarray}
Z_j (I)(t) = E\left(X_j(t)|(f_{i_1}(X_j),\ldots,f_{i_d}(X_j)\right),
\label{varunobservable}
\end{eqnarray}
and denote by $Q^*_n(I)$ it's empirical distribution.

\begin{eqnarray}
h_n(I) &=&\frac{1}{n}\sum_{j=1}^n\left(\int_a^b\alpha^n(t)\left(X_j(t)-Z_j(I)(t)\right)dt\right)^2 \label{hpcaempfa}+
\\&&\frac{1}{n}\sum_{j=1}^n\left(\int_a^b\alpha^n(t)\left(Z_j(I)(t)-\hat{X}_j(I)(t)\right)dt\right)^2 \label{hpcaempfb}+
\\&&\frac{2}{n}\sum_{j=1}^n\left(\int_a^b\alpha^n(t)\left(X_j(t)-Z_j(I)(t)\right)dt\right)\left(\int_a^b\alpha^n(t)\left(Z_j(I)(t)-\hat{X}_j(I)(t)\right)dt\right). \ \ \ \ \ \label{hpcaempfc}
\end{eqnarray}

First, we show that (\ref{hpcaempfa}) converges to (\ref{hpcafun}).
\begin{eqnarray}
&&\frac{1}{n}\sum_{j=1}^n\left(\int_a^b\alpha^n(t)\left(X_j(t)-Z_j(I)(t)\right)dt\right)^2 = \frac{1}{n}\sum_{j=1}^n\left(\int_a^b\left(\alpha^n(t)-\alpha(t)\right)\left(X_j(t)-Z_j(I)(t)\right)dt\right)^2 + \label{hpcaempfai} \\&&\frac{1}{n}\sum_{j=1}^n\left(\int_a^b\alpha(t)\left(X_j(t)-Z_j(I)(t)\right)dt\right)^2 +\label{hpcaempfaii}
\\&& \frac{2}{n}\sum_{j=1}^n\left(\int_a^b\left(\alpha^n(t)-\alpha(t)\right)\left(X_j(t)-Z_j(I)(t)\right)dt\right)\left(\int_a^b\alpha(t)\left(X_j(t)-Z_j(I)(t)\right)dt\right). \ \ \ \ \  \label{hpcaempfaiii}
\end{eqnarray}

By Cauchy-Schwarz's inequality we see that the right hand side of (\ref{hpcaempfai}) converges a.s. to zero,
\begin{eqnarray*}
&&\frac{1}{n}\sum_{j=1}^n\left(\int_a^b\left(\alpha^n(t)-\alpha(t)\right)\left(X_j(t)-Z_j(I)(t)\right)dt\right)^2 \leq
 \Vert \alpha^n - \alpha \Vert ^2_{L^2_{[a,b]}}  \frac{1}{n}\sum_{j=1}^n \Vert X_j-Z_j(I) \Vert^2_{L^2_{[a,b]}}.
\end{eqnarray*}

Dauxois et al. \cite{DPR82}, under mild regular conditions, show strong
consistency of the eigenvalues and their associated eigenvectors
(see Propositions 2 and 4 therein).
They establish that it is enough to show the convergence of the
covariance matrix in the operator space norm, assumption
\textbf{HP1} is required. More specifically, they prove that
if
$
\Vert \mathbf{\Gamma}_n-\mathbf{\Gamma}\Vert_{F}
\rightarrow 0 \mbox{ a.s., }
$ then
$\Vert \alpha^n_k - \alpha \Vert _{L^2_{[a,b]}}\rightarrow 0  \mbox{ a.s.,  } \mbox{ for all } 1\leq k\leq p,$
where $\alpha_k^n$ (respectively $\alpha_k$)
are the eigenvectors of $\mathbf{\Gamma}_n$, which is the empirical
covariance matrix associated with $P_n$ (respectively $\mathbf{\Gamma}$,
which is the covariance matrix associated with $P$).

Then $\Vert \alpha^n_k - \alpha \Vert _{L^2_{[a,b]}}$ converges a.s. to zero. By the Law of Large Numbers $\frac{1}{n}\sum_{j=1}^n \Vert X_j-Z_j(I) \Vert^2_{L^2_{[a,b]}}$ is bounded, even more, since
$\left\{\Vert X_j-Z_j(I)\Vert, \mbox{ for } j=1,\dots,n\right\}$ are iid random
variables with finite second moment (assumption \textbf{H2}), we have that,
$$\frac{1}{n}
\sum_{j=1}^n \Vert X_j-Z_j(I)\Vert_{L^2_{[a,b]}}^2 \rightarrow
E_P\left(\Vert X-Z(I)\Vert^2\right) \ \  \mbox{a.s.}$$

We will prove that (\ref{hpcaempfaii}) converges to (\ref{hpcafun}). By the Law of Large Numbers, given that

$\left\{\int_0^1 \alpha(t) (X_j(t)-Z_j(I)(t))dt, \mbox{ for } j=1,\dots,n\right\}$ are iid random
variables with finite second moment. Assumption \textbf{H2} and Cauchy-Schwarz's inequality ensures this statement since $$\left(\int_0^1 \alpha(t) (X_j(t)-Z_j(I)(t))dt\right)^2 \leq \Vert \alpha \Vert_{L^2_{[a,b]}} ^2  \Vert X_j-Z_j(I)\Vert_{L^2_{[a,b]}} ^2,$$  we have that,
$$\frac{1}{n}
\sum_{j=1}^n \left(\int_a^b \alpha(t) (X_j(t)-Z_j(I)(t))dt\right)^2 \rightarrow
E\left(\left(\int_a^b \alpha(t) (X(t)-Z(I)(t))dt\right)^2\right) \ \  \mbox{a.s.}$$
By Cauchy- Schwarz's inequality it can be seen that  (\ref{hpcaempfaiii}) and (\ref{hpcaempfc})  converges to zero.
An also (\ref{hpcaempfb}) vanishes by Cauchy-Schwarz's inequality and assumption \textbf{H1.}
%
\end{proof}

\begin{proof} \textbf{Theorem \ref{consistenciareg1}}

We need to proof that
\begin{eqnarray*}
&&\frac{1}{n}\sum_{j=1}^n \left(\int_a^b \beta_n(t)X_j(t)dt - \int_a^b
\beta_n(t)\hat{X}_j(I)(t)dt\right)^2 \rightarrow  E\left( \left( \int_a^b \beta_0(t)X(t)dt - \int_a^b
\beta_0(t)Z(I)(t)dt\right)^2\right) \ a.s.
\end{eqnarray*}

We consider the unobservable functions defined in (\ref{varunobservable}) and observe that
\begin{eqnarray}
h_n(I)&=& \frac{1}{n}\sum_{j=1}^n \left(\int_a^b \beta_n(t)X_j(t)dt - \int_a^b
\beta_n(t)\hat{X}_j(I)(t)dt\right)^2 \nonumber
\\&=& \frac{1}{n}\sum_{j=1}^n \left(\int_a^b \beta_n(t)\left(X_j(t)-Z_j(I)(t)\right)dt\right)^2 + \label{hnlinealfun1a}
\\&& \frac{1}{n}\sum_{j=1}^n \left(\int_a^b \beta_n(t)\left(Z_j(I)(t)-\hat{X}_j(I)(t)\right)dt\right)^2 + \label{hnlinealfun1b}
\\&& \frac{2}{n}\sum_{j=1}^n \left(\int_a^b \beta_n(t)\left(X_j(t)-Z_j(I)(t)\right)dt\right)\left(\int_a^b \beta_n(t)\left(Z_j(I)(t)-\hat{X}_j(I)(t)\right)dt\right). \ \ \ \ \ \label{hnlinealfun1c}
\end{eqnarray}

First, we are going to proof that  (\ref{hnlinealfun1a}) converges to $h(I)$. It is clear that (\ref{hnlinealfun1a}),
\begin{eqnarray}
&& \frac{1}{n}\sum_{j=1}^n \left(\int_a^b \beta_n(t)\left(X_j(t)-Z_j(I)(t)\right)dt\right)^2 = \frac{1}{n}\sum_{j=1}^n \left(\int_a^b \left(\beta_n(t)-\beta_0(t)\right)\left(X_j(t)-Z_j(I)(t)\right)dt\right)^2 + \label{hnlinealfun1ai}
\\&& \frac{1}{n}\sum_{j=1}^n \left(\int_a^b \beta_0(t)\left(X_j(t)-Z_j(I)(t)\right)dt\right)^2 + \label{hnlinealfun1aii}
\\&&  \frac{2}{n}\sum_{j=1}^n \left(\int_a^b \left(\beta_n(t)-\beta_0(t)\right)\left(X_j(t)-Z_j(I)(t)\right)dt\right)\left(\int_a^b \beta_0(t)\left(X_j(t)-Z_j(I)(t)\right)dt\right). \ \ \ \ \ \label{hnlinealfun1aiii}
\end{eqnarray}

We will show that the right hand side of (\ref{hnlinealfun1ai}) converges to zero. By  Cauchy-Schwarz's inequality, we have that,
\begin{eqnarray*}
\frac{1}{n}\sum_{j=1}^n \left(\int_a^b \left(\beta_n(t)-\beta_0(t)\right)\left(X_j(t)-Z_j(I)(t)\right)dt\right)^2
\leq \left\|\beta_n-\beta_0\right\|^2_{L^2[a,b]} \frac{1}{n}\sum_{j=1}^n \left\| X_j-Z_j(I) \right\|^2_{L^2[a,b]}.
\end{eqnarray*}
Since $\left\|\beta_n-\beta_0\right\|^2_{L^2[a,b]}$ converges a.s. to zero by assumption \textbf{HR1} and by the Law of Large Numbers we have that $\frac{1}{n}\sum_{j=1}^n \left\| X_j-Z_j(I) \right\|^2_{L^2[a,b]}$ is finite, because
$\left\{\left\| X_j-Z_j(I) \right\|^2_{L^2[a,b]} \text{ for } j=1,\ldots,n \right\}$ are iid random variables with finite second moment (assumption \textbf{H2}).

We will show that  (\ref{hnlinealfun1aii}) converges to $h(I)$. By Cauchy-Schwarz's inequality we have,
\begin{eqnarray*}
\left(\int_a^b \beta_0(t)\left(X(t)-Z(I)(t)\right)dt\right)^2 \leq \left\| \beta_0 \right\|^2_{L^2[a,b]} \left\| X-Z(I)\right\|^2_{L^2[a,b]},
\end{eqnarray*}
then,
\begin{eqnarray*}
E\left(\left(\int_a^b \beta_0(t)\left(X(t)-Z(I)(t)\right)dt\right)^2\right) \leq \left\| \beta_0 \right\|^2_{L^2[a,b]} E\left(\left\| X-Z(I)\right\|^2_{L^2[a,b]}\right).
\end{eqnarray*}
Given  that  $\left\{\int_a^b \beta_0(t)\left(X_j(t)-Z_j(I)(t)\right)dt \text{ for } j=1,\ldots,n \right\}$ are iid random variables with finite second
 moment (assumption \textbf{H2}), by the Law of Large Numbers we have that,
\begin{eqnarray*}
\frac{1}{n}\sum_{j=1}^n \left(\int_a^b \beta_0(t)\left(X_j(t)-Z_j(I)(t)\right)dt\right)^2 \rightarrow  E_P\left( \left( \int_a^b \beta_0(t)X(t)dt - \int_a^b \beta_0(t)Z(I)(t)dt\right)^2\right) = h(I) \ a.s.
\end{eqnarray*}

By Cauchy-Schwarz's inequality it can be seen that (\ref{hnlinealfun1c}), (\ref{hnlinealfun1b}) and (\ref{hnlinealfun1aiii}) converge a.s. to zero.

\end{proof}


\begin{proof} (\textbf{Theorem \ref{consistencialineal2fun}})
We need to proof that $h_n(I)\rightarrow h_(I)$ a.s., that means that,
\begin{eqnarray*}
\frac{1}{n}\sum_{j=1}^n\left[\int_c^d\left[\int_a^b\beta_n(s,t)(X_j(s)ds-\hat{X}_j(I)(s))ds\right]^2dt\right] \rightarrow E\left[\int_c^d\left[\int_a^b\beta(s,t)(X(s)-Z(I)(s))ds \right]^2dt\right] a.s.
\end{eqnarray*}

Observe that
\begin{eqnarray}
h_n(I)=\frac{1}{n}\sum_{j=1}^n\left[\int_c^d\left[\int_a^b\beta_n(s,t)(X_j(s)ds-\hat{X}_j(I)(s))ds\right]^2dt\right] \nonumber
\\= \frac{1}{n}\sum_{j=1}^n\left[\int_c^d\left[\int_a^b\beta_n(s,t)(X_j(s)ds-Z_j(I)(s))ds\right]^2dt\right] + \label{hnreg2conv1}
\\  \frac{1}{n}\sum_{j=1}^n\left[\int_c^d\left[\int_a^b\beta_n(s,t)(Z_j(I)(s)ds-\hat{X}_j(I)(s))ds\right]^2dt\right] + \label{hnreg2conv2}
\\\frac{2}{n}\sum_{j=1}^n\left[\int_c^d\left[\int_a^b\beta_n(s,t)(X_j(s)ds-Z_j(I)(s))ds\right]\left[\int_a^b\beta_n(s,t)(Z_j(s)ds-\hat{X}_j(I)(s))ds\right]dt\right].\label{hnreg2conv3}
\end{eqnarray}

First we are going to see that (\ref{hnreg2conv1}) converges to $h(I)$
\begin{eqnarray}
\frac{1}{n}\sum_{j=1}^n\int_c^d\left[\int_a^b\beta_n(s,t)(X_j(s)ds-Z_j(I)(s))ds\right]^2dt
= \frac{1}{n}\sum_{j=1}^n\int_c^d\left[\int_a^b\beta(s,t)(X_j(s)ds-Z_j(I)(s))ds\right]^2dt \label{hnreg2conv1a}
\\+ \frac{1}{n}\sum_{j=1}^n\int_c^d\left[\int_a^b(\beta_n(s,t)-\beta(s,t))(X_j(s)ds-Z_j(I)(s))ds\right]^2dt \label{hnreg2conv1b}
\\+ \frac{2}{n}\sum_{j=1}^n\int_c^d\left[\int_a^b\beta(s,t)(X_j(s)ds-Z_j(I)(s))ds\right]
 \left[\int_a^b(\beta_n(s,t)-\beta(s,t))(X_j(s)ds-Z_j(I)(s))ds\right]dt \label{hnreg2conv1c}
\end{eqnarray}

We are going to proof that (\ref{hnreg2conv1a}) converges to $h(I)$. Since $\int_c^d\left[\int_a^b\beta(s,t)(X_j(s)ds-Z_j(I)(s))ds\right]^2 dt$ has finite first moment, by Cauchy-Schwarz's inequality we have that
\begin{eqnarray*}
\int_c^d\left[\int_a^b\beta(s,t)(X_j(s)ds-Z_j(I)(s))ds\right]^2 dt
\leq \Vert X_j-Z_j(I) \Vert^2_{L^2[a,b]} \Vert\beta\Vert^2_{L^2([a,b]\times[c,d])}.
\end{eqnarray*}

Then
\begin{eqnarray*}
E\left(\int_c^d\left[\int_a^b\beta(s,t)(X_j(s)ds-Z_j(I)(s))ds\right]^2 dt\right)\leq  \Vert\beta\Vert^2_{L^2([a,b]\times[c,d])} E(\Vert X_j-Z_j(I) \Vert^2_{L^2[a,b]}),
\end{eqnarray*}
which is finite by assumption \textbf{H2}.

Since $\left\{\int_c^d\left[\int_a^b\beta(s,t)(X_j(s)ds-Z_j(I)(s))ds\right]^2 dt\right\}$ are iid variables with finite first moment, by Law of the Large Numbers, we have that (\ref{hnreg2conv1a}) converge a.s. to (\ref{hlinealfun2}).

Assumptions \textbf{H2} and \textbf{HRF1}, and Cauchy-Schwarz's inequality guarantee that (\ref{hnreg2conv1b}) converges a.s. to zero.

As a consequence of Cauchy-Schwarz's inequality it can be seen that (\ref{hnreg2conv1c}) converges a.s. to zero.

As a consequence of assumptions \textbf{H2} and \textbf{HRf1}, and by Cauchy-Schwarz's inequality it can be seen that (\ref{hnreg2conv2}) converges a.s. to zero.
Finally it can be seen that that assumptions \textbf{H1}, \textbf{H2} and \textbf{HRF1}, and Cauchy-Schwarz's inequality guaranty that (\ref{hnreg2conv3}) converges a.s. to zero.

\end{proof}

\end{document}